\documentclass[%
 aip,
 amsmath,amssymb,
 reprint,%
]{revtex4-1}

\usepackage{graphicx}
\usepackage{dcolumn}
\usepackage{bm}

\usepackage[utf8]{inputenc}
\usepackage[T1]{fontenc}
\usepackage{mathptmx}
\usepackage{etoolbox}

\makeatletter
\def\@email#1#2{%
 \endgroup
 \patchcmd{\titleblock@produce}
  {\frontmatter@RRAPformat}
  {\frontmatter@RRAPformat{\produce@RRAP{*#1\href{mailto:#2}{#2}}}\frontmatter@RRAPformat}
  {}{}
}%
\makeatother
\begin{document}

\preprint{AIP/123-QED}

\title[Jupiter's Vortex Crystals]{Jupiter's Polar Vortex Crystals 
Explored using the Shallow Water Equations}

\author{Sihe Chen}
 \email{sihechen@caltech.edu.}
 \affiliation{Planetary Science Department, California Institute of Technology.}
\author{Andrew P. Ingersoll}%
\affiliation{Planetary Science Department, California Institute of Technology.}%

\author{Cheng Li}
\affiliation{Department of Climate and Space Sciences and Engineering, University of Michigan, Ann Arbor.}%

\date{\today}

\begin{abstract}
At the poles of Jupiter, cyclonic vortices are clustered together in patterns made up of equilateral triangles called vortex crystals. Such patterns are seen in laboratory flows but never before in a planetary atmosphere, where the planet's rotation and gravity add new physics. Here we use the shallow water (SW) equations at the pole of a rotating planet to study the emergence and evolution of vortices starting from an initial random pattern of small-scale turbulence. The flow is in a single layer with a free surface whose slope produces the horizontal pressure gradient force. We explored three parameters in the problem: the mean kinetic energy of the initial turbulence, the horizontal scale of the initial turbulence, and the radius of deformation of the undisturbed fluid layer. We find that some regions of this parameter space lead to vortex crystals and others lead to chaotic behavior and mergers. Our results identified that the relative change of the layer thickness is the key quantity that determines whether the vortex crystal or chaotic patterns emerge. In doing so we are learning about Jupiter's atmosphere – the small-scale turbulence and the vertical structure of the layer in which the cyclones exist. We are also learning about basic physics since vortex crystals are a fundamental phenomenon in fluid mechanics.
\end{abstract}

\maketitle

\section{Introduction}
At each pole of Jupiter, a central cyclonic vortex is surrounded by a ring of cyclonic vortices at a latitude of ~83°,  corresponding to a circle  ~8700 km in radius. The individual cyclones have radii, as defined by their azimuthal velocity maxima, of ~1000 km. In the north, the ring has 8 circumpolar cyclones, and in the south, the ring has 5 circumpolar cyclones. Thus the central cyclone and the cyclones around it resemble approximately a two-dimensional structure made up of equilateral triangles. Such patterns have been studied in the laboratory and are appropriately called vortex crystals \cite{Fine1995,Schecter1999,Siegelman2022b}.

Here we use the shallow water (SW) equations at the pole of a rotating planet to study the emergence and evolution of vortices starting from an initial random pattern of small-scale turbulence. The flow is in a single layer, of thickness h, with a free surface that is capable of propagating gravity waves. We use a reduced-gravity model\cite{Chassignet1991}, where the active layer is floating hydrostatically on a much deeper fluid layer at rest. We use the reduced gravity $g_r=g\Delta\rho/\rho$, a constant, where $g$ is the gravitational acceleration at the planet's surface and $\Delta\rho/\rho$ is the fractional density difference between the two layers. Then the geopotential, $\Phi=g_rh$, is one of the variables of the SW system, the other being the velocity $\mathbf{u}$. The upper boundary of the active layer is a free surface with zero pressure, so $-\nabla\Phi$ is the horizontal pressure gradient force per unit mass within the layer. Since $\Phi$ is proportional to the thickness, it obeys the horizontal continuity equation. This, along with the horizontal momentum equation, the full set of equations of the SW system is constituted:
\begin{eqnarray}
    \frac{\partial\Phi}{\partial t}+(\mathbf{u}\cdot\nabla)\Phi+\Phi\nabla\cdot\mathbf{u}=0\label{eq:eq1}\\
    \frac{\partial\mathbf{u}}{\partial t}+(\mathbf{u}\cdot\nabla)\mathbf{u}+f\hat{\mathbf{k}}\times\mathbf{u}=-\nabla\Phi\label{eq:eq2}
\end{eqnarray}
We are modeling a thin horizontal layer on a rotating sphere, so the vertical component of the Coriolis acceleration is what matters. Thus $\hat{\mathbf{k}}$ is the vertical unit vector and $f=2\Omega\sin\phi$ is the Coriolis parameter, where $\Omega$ is the planetary rotation rate and $\phi$ is latitude. In this paper, we use 3 parameters – the root mean square velocity $U$ of the initial turbulence, the horizontal wavelength $L_i$ of the initial turbulence, and the radius of deformation $L_d$ of the undisturbed fluid layer. The planetary radius $a$ is 71,492 km, and the period of rotation is 9$^h$ 55$^m$ 29$^s$. These two numbers do not change, and serve to define the units of length and time, respectively. Our results hold for other planets when the units are re-scaled accordingly. Other approaches use planet size as an additional parameter\cite{BRUESHABER2019}, but our approach keeps the number of parameters to a minimum. We explore this 3D parameter space and find that some regions lead to vortex crystals and others lead to chaotic behavior and mergers.

One might ask if this single-layer SW model is realistic enough to capture all the important parameters of Jupiter’s atmosphere: the nature of the small-scale turbulence, e.g.,  whether it arises from convection, baroclinic instability, breaking gravity waves, or breaking planetary waves; the thickness and vertical structure of the layer where the cyclones exist; the nature of dissipation and whether there are upscale and downscale cascades of energy and so on. The answer is no, but we cannot achieve perfect realism. There is no sufficient observational data for the poles, especially about vertical structure. The current knowledge is related to small-scale turbulence and inverse cascades, at least near cloud-top altitudes\cite{Ingersoll2022,Siegelman2022a}. In the future, we are gaining information from the Juno mission about the roots of the vortices and stable layers at depth. This paper explores the range of parameters that produce vortex crystals, providing insights to such measurements, and constraints to the future modeling work of the polar region.

\section{Dimensionless numbers and length scales}
Dimensionless numbers are useful in estimating the relative importance of different terms in the equations. The Rossby number $U/(fL_i)$, where $f=2\Omega$ is the Coriolis parameter at the poles. It gives the importance of fluid accelerations relative to the Coriolis acceleration, and the former arises from motions on the scale of the initial disturbance. The individual cyclones at Jupiter’s poles have peak azimuthal velocities of $~$80 m/s and peak relative vorticities $\zeta=\hat{\mathbf{k}}\cdot(\nabla\times\mathbf{u})\approx3.0 \times 10-4 $s$^{-1}$, the latter being almost equal to the planetary vorticity $f$ at the pole of $3.5 \times 10^{-4 }$ s$^{-1}$ \cite{Ingersoll2022}. The alternative definition for the Rossby number is $\zeta/f$, which is approximately 1 for the large cyclones and may be larger than 1 for the initial turbulence.

The reduced gravity $g_r$ and layer thickness $h$ appear in the equations only as a single product, which is the variable $\Phi = g_rh$ and has dimensions of velocity squared. The global mean value of $\Phi$ divided by $f^2$ gives $L_d^2$, the square of the radius of deformation. It measures the mean thickness and vertical structure of the fluid layer and is the second free parameter that we vary. Large $L_d$ is related to vertical stability. For a given velocity, large $L_d$ means that pressure surfaces are nearly horizontal and the horizontal flow is nearly incompressible, as seen in Eq. (\ref{eq:eq1}).

The third parameter that we vary is the horizontal scale $L_i$ of the initial small-scale turbulence. As detailed in the next section, this parameter determines the wavelengths of the initial turbulence, which is the lengthscale of which the energy cascade starts. The Rossby number $U/(fL_i)$ of the initial field in our experiments then ranges from 0.1 to 2.3.

The initial $dh/h$ is a combination of the three basic initial parameters $U$, $L_d$, and $L_i$. It is the fractional departure of the mean surface height from equilibrium, and when the value becomes of order 1, the shallow water thickness can potentially become negative. This causes the sound speed and hence the Courant–Friedrichs–Lewy (CFL) number to become undefined, and we stop the simulation when this happens. Physically, this does not mean a complete failure of the shallow water system: it could be trying to represent breaking gravity waves in the 3D physical system. The quasigeostrophic (QG) equations have gravity waves filtered out and are based on the assumptions of small perturbation to the thickness and small Rossby number: $dh/h \ll 1$ and $U/(fL_i) \ll 1$. So the SW equations are more general than QG. To estimate the value of $dh/h$ in the SW system, we use a scaling based on cyclostrophic balance:
\begin{equation}
    \frac{u^2}{r}+fu=-g_r dhdr,
\end{equation} 
Taking $v \sim U$, $dr \sim L_i$, and $L_d^2 \sim g_r h$, we obtain 
\begin{equation}
    dh/h \sim \frac{U^2+fUL_i}{(fL_d)^2}.\label{eq:dhh}
\end{equation} 
This covers both small and large values of the Rossby number $U/(fL_i)$. When the Rossby number is small, the initial scale of the turbulence matters, which gives
\begin{equation}
    dh/h \sim \frac{fUL_i}{(fL_d)^2};
\end{equation} 
When the Rossby number becomes large, the velocity dominates, and the initial scale of turbulence has minimal influence on the thickness perturbation:
\begin{equation}
    dh/h \sim \frac{U^2}{(fL_d)^2}.
\end{equation} 
In this paper, we use the expression presented in Eq. (\ref{eq:dhh}) as to calculate the values of $dh/h$.

The increase of $f$ with latitude – the beta effect, works differently at the poles than at mid 
latitudes. Here $\beta=df/⁄dy=2\Omega \cos\phi/a$, where $y$ is the poleward coordinate and $a$ is the radius of the planet. At the mid-latitudes, the quantity $\sqrt{U/\beta}$ is a length scale below which eddies dominate and above which zonal jets dominate \cite{Rhines1975}. The turbulence is assumed to be cascading from small scales to large scales, but when it reaches the scale of $\sqrt{U/\beta}$, the cascade stops. However, since $\beta$ goes to zero at the poles, the cascade can never reach that scale at the poles, and eddies should always dominate. 

The question then arises: at what latitude does the transition, with zonal jets below that latitude and eddies above it, take place? If we expand $\beta$ at radial distance $r$ away from the pole using its derivative at the pole: $\beta\approx 2\Omega r/a^2$, and assume that the eddy scale inside $r$ is $r$ itself, as on Jupiter. Then 
\begin{equation}
    \sqrt{\frac{U}{\beta}}\approx \sqrt{\frac{Ua^2}{2\Omega r}}\approx r.
\end{equation} 
Solving for $r$ gives
\begin{equation}
    r \approx\left(\frac{Ua^2}{2\Omega}\right)^{\frac{1}{3}}. \label{eq:Lgamma}
\end{equation}
The answer has been given before \cite{Ingersoll2022,Siegelman2022b}, but here we applied an assumption about the eddies to provide the physical intuition of the lengthscale. With U = 80 m/s, the expression gives 10,500 km, which is a fairly good match to the 8700 km radius of the circumpolar ring. Note that this critical latitude varies as $U^{1/3}$, and it is only a scaling estimate, not a quantitative theory.

\section{Model description}
\begin{figure}
\includegraphics[width=0.5\textwidth]{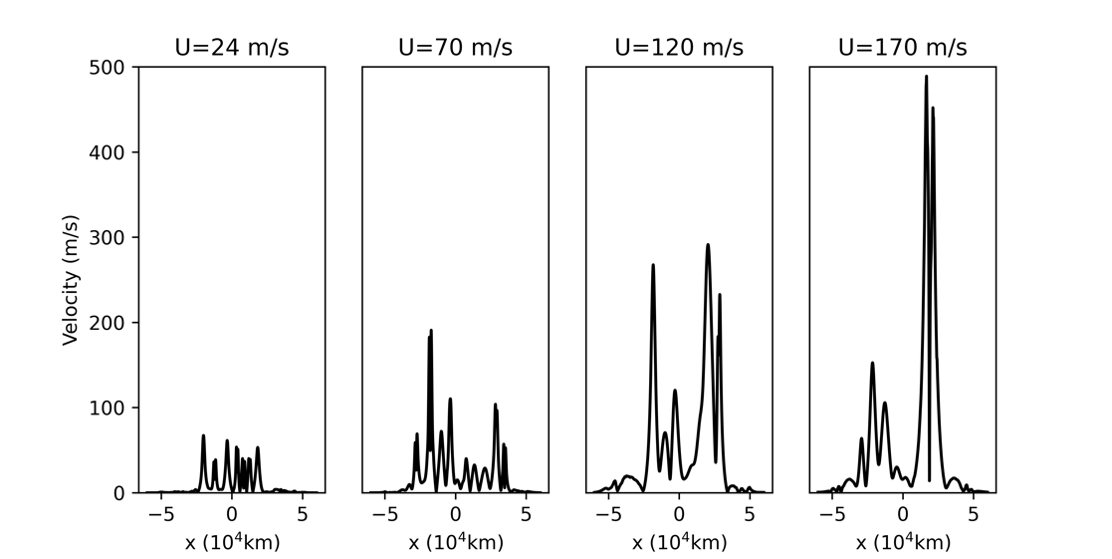}
\caption{\label{fig:sample_vel} Velocity magnitudes along a central chord of the computational domain for four different values of $U$, all with $L_i=$350 km and $L_d=$2100 km.}
\end{figure}

We solve shallow-water equations using an Athena++-based shallow-water model \cite{Li2019,Stone2020} to study the genesis of the vortices and their subsequent evolution. 
The setup of the initial condition is similar to Siegelman et al. (2022)\cite{Siegelman2022b}. We assume that the initial streamfunction is a superposition of sinusoidal waves. For each wave numbered $n$, it has an argument of $k(x \cos \theta_n+y \sin \theta_n  + \alpha_n)$. Here k is the wavenumber, within a $\pm5 \%$ range from the central wavenumber specified by $2\pi/L_i$. $\theta_n$ and $\alpha_n$ are random numbers between $0$ and $2\pi$, meaning that the waves are from a narrow ring in wavenumber space and uniformly distributed in direction. The initial velocity field is geostrophically balanced and is populated everywhere within a region extending from the pole to a distance 5$\times$10$^4$ km away from the pole. This region extends down to 50° of latitude, making it much larger than the vortices themselves. The amplitude of the noise is chosen so that the root mean square velocity in the field between 60 degrees and the pole is $U$. The choice of 60 degrees is made as most of the energy in the final patterns is always confined within this region. The final magnitudes of velocities observed in the final patterns, corresponding to each of the initial U selections, are shown in Figure \ref{fig:sample_vel}.

The jovian vortices were measured to have velocity profiles that reach 80 to 110 m/s at approximately 1000 km from their centers \cite{Grassi2018}. This implies that, for $U$ values of 24 m/s and 70 m/s, the velocity patterns are closer to the observations. Hence, we chose U values of 24, 70, 120, and 170 m/s, bracketing and exploring beyond the observed values. The value of $L_d$ in the polar region is not measured, but the mid-latitude value is estimated to be 1000 km \cite{WONG2011}. Various estimates for $L_d$ in the polar region span the range from 300 km to 1300 km. For the central vortex in the north, the requirement that the thickness cannot be negative leads to the smallest value of $L_d$ being at least 750 km \cite{Ingersoll2022}. Hence, we choose $L_d$ values to be well larger than this lower bound, ranging from 1750 to 8400 km. For $L_i$, we choose turbulence scales that are smaller than the scale of the vortices, i.e., 210, 350, and 700 km. For the choice of the parameters, we focused on regions of transition between an unstructured pattern and vortex crystals, which leads to different $L_d$ choices with different $L_i$ values.

The flow takes place in a plane tangent to the surface of the planet at the pole. For most of the scenarios, we choose a 2048$\times$2048 grid, with $x_{max}=y_{max}=6\times10^4$  km, and $x_{min}=-x_{max}, y_{min}=-y_{max}$. This domain covers from the pole to about 40 degrees of latitude. With a grid convergence test, we found that $\Delta x\leq L_i/2$ is necessary to resolve the initial turbulence. For an initial forcing scale of 105 km, a model with doubled resolution has to be used, increasing the computational burden by 8 times. Hence, we start our exploration from the minimum length scale Li=210 km.

Additionally, we impose a sponge layer that starts at radius $r_*=5\times10^4$ km from the pole, corresponding to 50 degrees of latitude. The velocities are dissipated in the sponge layer by Rayleigh friction. The dissipation rate increases linearly from $0$ at $r = r_*$ to $0.5$ day$^{-1}$ at $r = 1.1 r_*$.

\section{Results}
\begin{figure}
\includegraphics[width=0.5\textwidth]{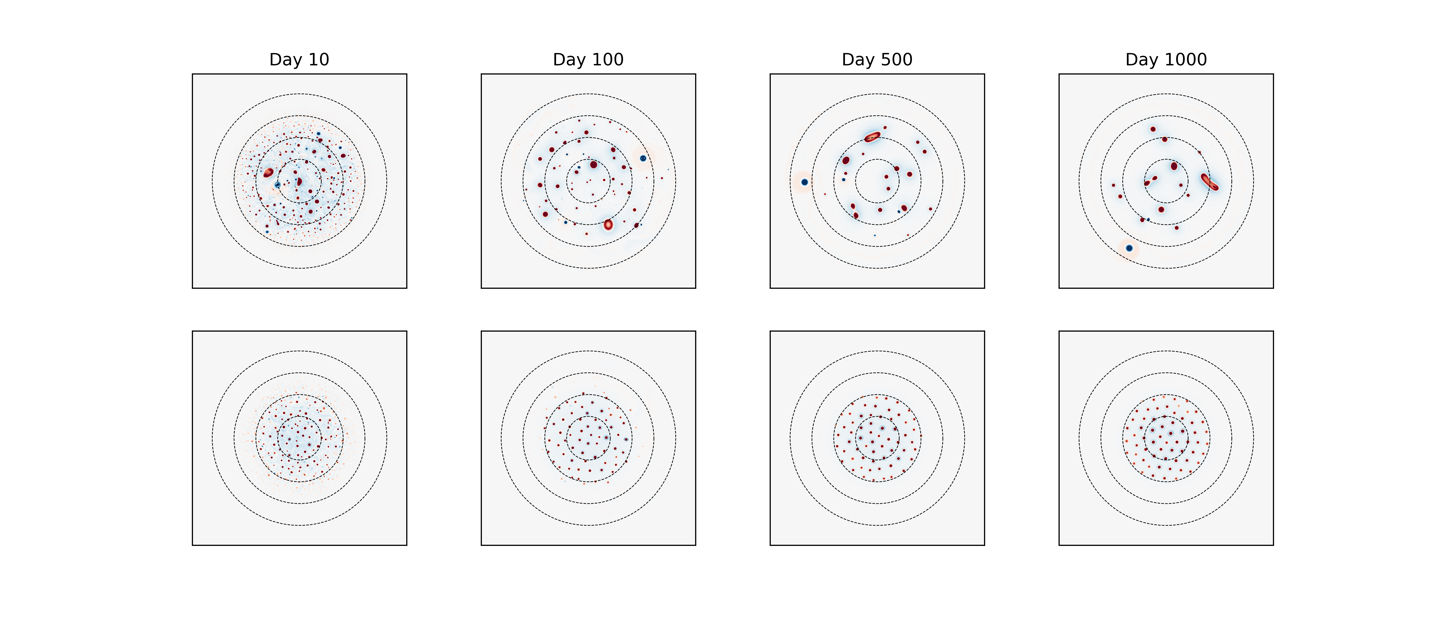}
\caption{The vorticity contours of two cases: an example where the vortices do not form a regular pattern, and an example where the vortices organize to a vortex crystal. The patterns at days 10, 100, 500, and 1000 are presented. The circles are at latitudes of 80, 70, 60, and 50 degrees. Parameter choices for each of the two cases: Case 1 (upper row): $L_i=$210 km, $U=$120 m/s, and $L_d=$2100 km; Case 2 (lower row): $L_i=$350 km, $U$=70 m/s, and $L_d$=7000 km. These values give large $dh/h=0.043$ for Case 1 and small $dh/h=0.002$ for Case 2, respectively.}
\label{fig:time_comp}
\end{figure}

Fig. \ref{fig:time_comp} shows flow undergoing an inverse energy cascade. The small-scale energy contained in the initial turbulence cascades into larger scales, forming structured vorticity. The energy thus flows from kinetic energy to potential energy, while the scale at which the energy cascade stops varies by case. In the first 10 days, distinct behavior starts to be present. Both scenarios develop vortices from the initial turbulence within days after the start of simulation. However, even though Case 1 has a smaller initial forcing scale, it ends up with larger vortices, which means more extensive merging takes place. In comparison, Case 2 shows a regular pattern of similarly sized vortices. The following development of the patterns are also different: Case 1 continues to merging and forming larger vortices, while Case 2 stop merging and settles in to a vortex crystal configuration, not showing significant changes in pattern for over a thousand days.

\begin{figure*}
    \includegraphics[width=\textwidth]{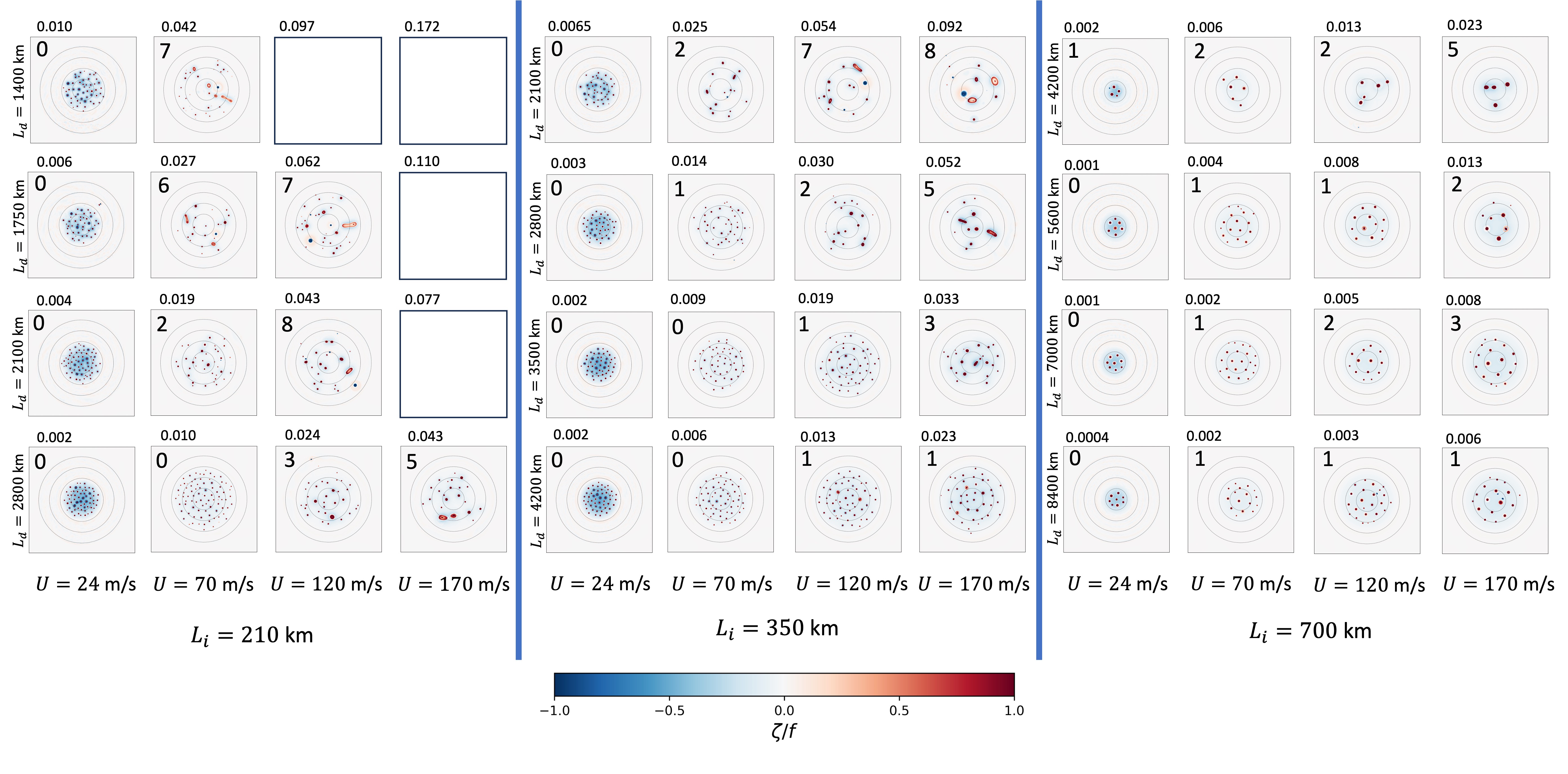}
    \caption{The vortex patterns at day 600. The full dataset simulated is shown, and values of $dh/h$ are presented as the number on top of each frame. The numbers placed within the frames in the top-left corners mark the level of irregularity based on the $dA/A$ metric described in the text. Note that the $L_d$ values along the y-axis are different for the three panels. This change is to focus on the transition from irregular patterns with relatively small $L_d$ values to vortex crystal patterns with large $L_d$ values. The value of this transition, between $dA/A \leq 1$ and $dA/A>1$, also varies with U and Li, which is why the axes are different.}
    \label{fig:vort_pattern}

\end{figure*}

In Fig. \ref{fig:vort_pattern}, we organize 48 images at day 600 and label them by $dh/h$. With a glance, it becomes apparent that cases with small $dh/h$ show a more regular pattern, closer to a vortex crystal, as compared to cases with large $dh/h$. Each image has a pattern, and we want a simple but quantitative way to classify it. We start by locating all the vortices in the image - patches of fluid whose vorticity is larger than $0.1f$, regardless of the sign. Then we calculate the area of each patch and the mean area $A$ and standard deviation $dA$ of all the patches in the image. The metric measuring the irregularity of the image is then $dA/A$. For aid in graphical depiction, we put the measured values of all the $dA/A$ values, one for each image, into 9 bins from low to high. We number the bins with single digits from 0 to 8.

The levels of irregularity defined as the single-digit numbers in Fig. \ref{fig:vort_pattern} agree well with our subjective impression. Numbers less than 2 look like regular vortex crystals. Numbers 2 to 5 look somewhat irregular, and 6 to 8 look chaotic and irregular. It is not a perfect metric, and the connection to $L_d$ and other input parameters is not perfect, perhaps because there are random elements in the merging process. Blank spaces in the left panel are where $dh/h$ was so large that the program crashed before reaching 600 days. Crashing was always due to the geopotential $\Phi$ becoming negative.

If one scans across any of the 4 images with constant $L_d$ and $L_i$, it is clear that irregularity increases with $U$. Similarly, if one scans down any of the 4 images with constant $U$ and $L_i$, it is clear that irregularity decreases as $L_d$ increases. Both of these observations are consistent with irregularity increasing with $dh/h$, which has $fUL_i + U^22$ in the numerator and $L_d^2$ in the denominator. As shown in Fig. \ref{fig:time_comp}, for smaller $dh/h$ values, the patterns stop evolving after the initial turbulence is organized into vortices. However, for larger $dh/h$ values, the pattern remains evolving for a long time, allowing the vortices to merge, which does not favor a vortex crystal structure.

\begin{figure*}
    \includegraphics[width=\textwidth]{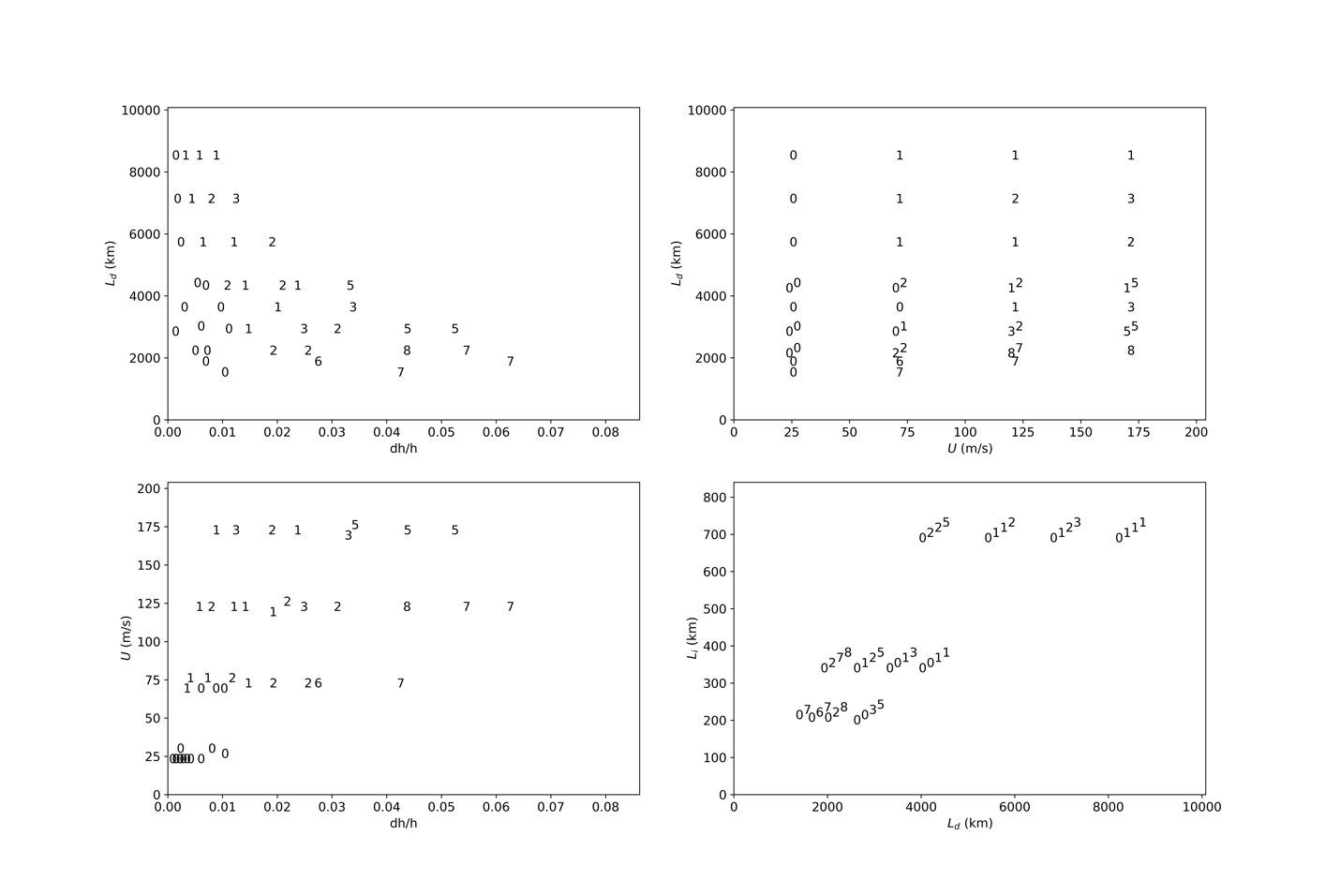}
    \caption{Levels of irregularity for each image in Fig. \ref{fig:vort_pattern} as a function of input parameters. The irregularity of a case is its dA/A value expressed as a single-digit number from 0 to 8 as described in the text. The input parameters $U$, $L_d$, and $L_i$ define a 3D space, and the variable $dh/h$ is computed from those parameters. There are no double-digit numbers. Digits that share the same locations are offset by a small amount to avoid overlapping.
}
    \label{fig:Digit_irr}

\end{figure*}

Fig. \ref{fig:Digit_irr} is a regime diagram in the 3D space defined by the input parameters $U$, $L_d$, and $L_i$. The output variable is $dA/A$, which measures the computed patterns' irregularity. Vortex crystals are the most regular and have the lowest values of $dA/A$. Chaotic behavior and mergers have the highest values. In principle, $dA/A$ fills parameter space and summarizes the result of our investigation, and it is useful to study how the low, intermediate, and high values of $dA/A$ are distributed throughout parameter space after 600 days, after which the patterns have settled down. We do this by treating the parameter space as a cube and projecting the $dA/A$ values along the 3 axes. We added another input parameter, $dh/h$, which is a combination of the original three. It is found that the combined variable $dh/h$ is more closely related to vertical structure than the others. $L_d$, on the other hand, is more closely related to horizontal structure: it is the e-folding distance over which vortices' interaction with each other drops.

In the upper right panel of Fig. \ref{fig:Digit_irr}, irregularity bin numbers of 3 or more are confined to the lower right corner of the figure, where $U$ is large and $L_d$ is small, specifically $L_d\leq 4000$ km for $U = $170 m/s and $L_d \leq$2000 km for $U = 70$ m/s. This suggests that $dh/h = (U^2+fUL_i)/(fL_d )^2$ is the controlling variable. This is borne out by the two left panels of Fig. \ref{fig:Digit_irr}, which always show large $dA/A$ bins when $dh/h$ is large. Both large $U$ and small $L_d$ seem to be necessary for merging and chaos. Otherwise, vortex crystals are favored. From the definition of $dh/h$, it appears that $L_i$ has an effect only when the initial Rossby number is less than 1.

The upper left panel of Fig. \ref{fig:Digit_irr} has the integers representing $dA/A$ varying almost monotonically along a line in the projected parameter space. The small integers are in the upper left corner, and the large integers are in the lower right corner. The variables along the two axes are highly correlated and are responsible for most of the variations from regularity to irregularity from one run of the model to the next. This is an expected observation because the formula for $dh/h$ has $U$ and $L_i$ in the numerator and $L_d^2$ in the denominator. $L_i$ has a relatively small effect on the level of irregularity.

\begin{figure}
    \includegraphics[width=0.45\textwidth]{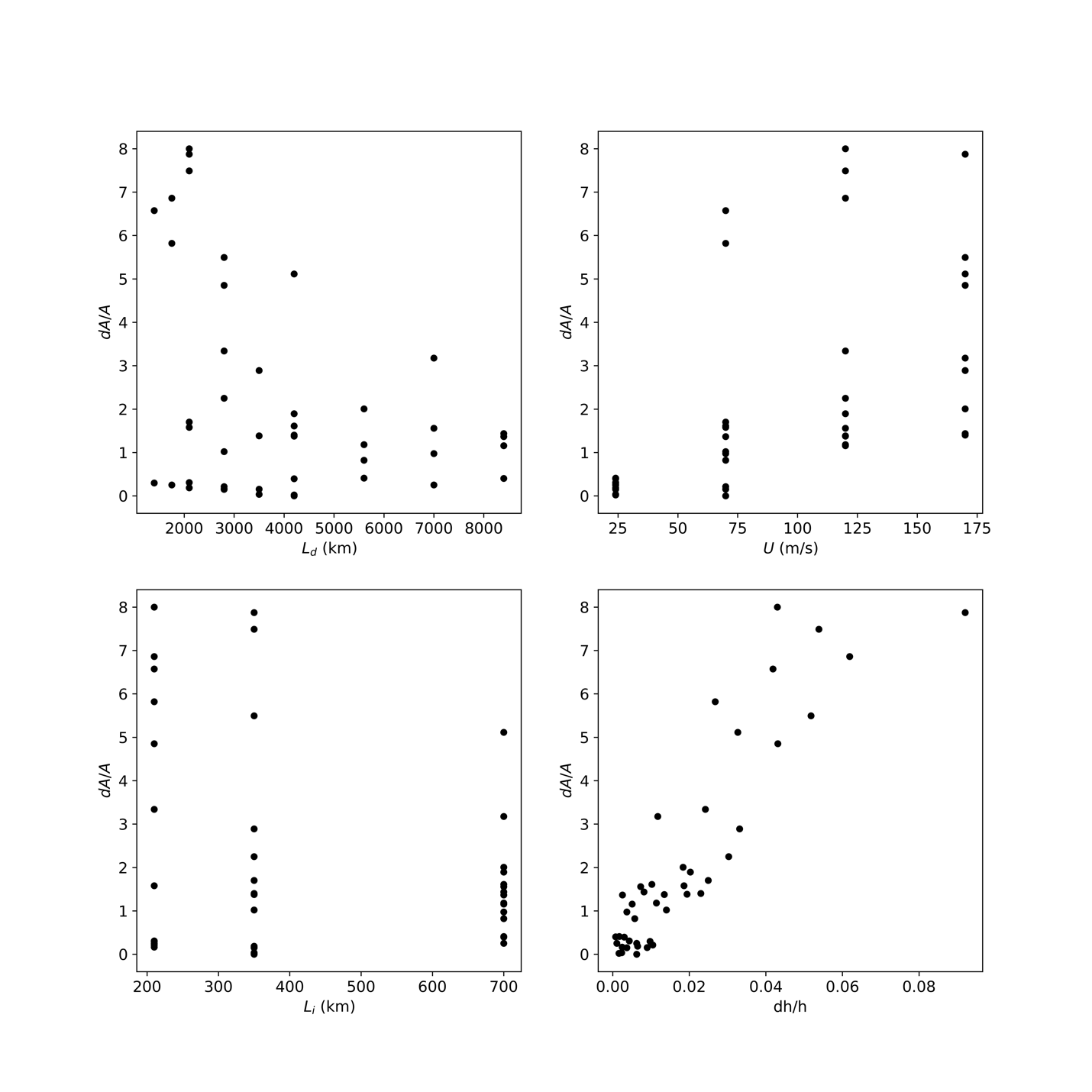}
    \caption{Scatter plots of variables versus the irregularity measure $dA/A$.
}
    \label{fig:Scatt_irr}

\end{figure}

A detailed view of the metric of irregularities is shown in Fig. \ref{fig:Scatt_irr}. While we did the parameter search by using various values of $L_d$, $L_i$, and $U$. We used $dh/h$ as a combined parameter that contains the information of all three variables. We obtain more insights on the effect of each parameter: the smaller radius of deformation $L_d$ shows a general trend resulting in a more irregular pattern. Larger characteristic velocity $U$ results in a slightly more irregular pattern.

While Fig. \ref{fig:Digit_irr} shows 2D landscapes of levels of irregularities, Fig. 5 shows the $dA/A$ bin numbers as scatter plots, with $L_d$, $U$, $L_i$, and $dh/h$ as the independent variables placed in the x-axis. As seen in the lower right panel, $dh/h$ does the best job of predicting when the flow will evolve to large $dA/A$, signifying an irregular pattern, as contrasted with small $dA/A$, which signifies a vortex crystal pattern. This is consistent with $dh/h$ being the controlling variable.

It has been noticed in Fig. \ref{fig:Scatt_irr} that there is a strong and linear correlation between the value of $dh/h$ to the level of irregularities, which is not obtained with the three original variables. This finding bridges the earlier results and discussions related to Jupiter vortices: models of shallow water did not create results with vortex crystal patterns; however, with a QG model, the generation of vortex crystal was obtained \cite{Siegelman2022b}. Hence, we suggest that having small $dh/h$, which is obtained when the atmosphere has low velocity and large vertical stability, helps the formation of stable vortex crystals, and that may be the case with Jupiter.

\section{Discussions \& Conclusions}
The perturbation to the layer thickness, $dh/h$, emerged as a dominant factor influencing the subsequent phases of vortex evolution. This phenomenon depicts the interplay between two-dimensional turbulence and the Coriolis effect in a rotating frame, which poses a limitation on the inverse energy cascade to larger scales.

We provide a qualitative argument that explains this phenomenon here. In the context of a non-rotating frame of reference, two-dimensional turbulence exhibits an inverse cascade behavior, transporting energy towards larger scales. In the absence of the Coriolis force, the inherent tendency of the vortex is to merge, culminating in the largest scale the system permits. 

However, the narrative shifts in a rotating frame, especially at the poles. In the rotating spheres, a beta effect is present that drives the cyclones to move along the positive potential vorticity gradient and anti-cyclones along the negative. Surrounding the geostrophically balanced cyclones, there are rings of negative vorticity, which form a negative potential vorticity gradient, shielding other vortices from moving in and merging. This is well observed, and such a structure is identified as the shielding of the vortex \cite{Li2020,Ingersoll2022}. The shielding effects change the dynamic behaviors, stopping further inverse cascades by inhibiting the merging of vortices into larger scales. This results in a cut-off, keeping the inverse energy cascade at relatively small scales.

This dynamic is encapsulated in Fig. \ref{fig:vort_pattern}, vortices with a larger $dh/h$ persist in channeling energy to larger scales, while their counterparts with a smaller $dh/h$ stopped this process, preserving a nearly constant scale, which eventually forms a vortex crystal configuration.

Another observation is associated with the radius of deformation. In previous publications, the vortex crystal is produced in an infinite-radius-of-deformation QG model\cite{Siegelman2022b}. In previous work that produced a single polar vortex instead of vortex crystals, the values of $dh/h$ are relatively large due to small values of radius of deformation (e.g. ~1700 km\cite{ONeil2015}, and ~1300 km \cite{BRUESHABER2019}). The shallow-water case where the vortex crystal stably exists, on the other hand, explored a radius of deformation up to 10,000 km. Hence, the underestimation of the polar radius of deformation may be the major reason why a vortex crystal was not predicted before being observed. The polar atmosphere of Jupiter may have a higher buoyancy frequency, i.e. may be more stably stratified than previously estimated.

Our results offer a tangible linkage from the strength and scale of forcing to the resulting vortex patterns: weak and small-scale perturbations result in small values of $dh/h$, more QG dynamics, and more regular vortex lattice patterns. On the other hand, when the perturbation is strong and large-scale, the values of $dh/h$ are large, and the vortices tend to merge and form larger structures, resulting in more Saturn-like polar patterns.

\begin{acknowledgments}
This research was carried out at the California Institute of Technology under a contract with the National Aeronautics and Space Administration (NASA), Grant/Cooperative Agreement Number 80NSSC20K0555, and a contract with the Juno mission, which is administered for NASA by the Southwest Research Institute. CL was supported by the NASA Juno Program, under NASA Contract NNM06AA75C from the Marshall Space Flight Center, through subcontract 699056KC to the University of Michigan from the Southwest Research Institute. We also thank Dr. Huazhi Ge for his suggestions in developing the simulations presented in this paper.
\end{acknowledgments}

\section*{Data Availability Statement}
The model program is made public on GitHub under the repository \url{https://github.com/chengcli/canoe}. The simulation output data that support the findings of this study are available from the corresponding author upon reasonable request.

\section*{bibliography}
\nocite{*}
\bibliography{aipsamp}

\end{document}